Integrals of motion on extremals of the equation Euler - Lagrange

V. P. Koshcheev

Moscow Aviation Institute (National Research University), Strela Branch, Zhukovsky, Moscow Region, Russia

email: koshcheev1@yandex.ru

Abstract: It is shown that a chain of closed systems of first-order ordinary differential equations describing the evolution of moments can be constructed using the Jacobi equation. It is shown that Wronsky determinants for fundamental matrices of closed systems of first–order ordinary differential equations are integrals of motion on extremals of the Euler-Lagrange equation. It is shown that the distribution function is localized in the vicinity of a classical trajectory, the size of which in the phase space is of the order of Planck's constant.

It was shown in [1] that a chain of closed systems of first-order ordinary differential equations describing the evolution of moments can be constructed using the Jacobi equation.

Let us be given the Euler–Lagrange equation

$$\frac{\partial L}{\partial x} - \frac{d}{dt}\frac{\partial L}{\partial \dot{x}} = 0, \qquad (1)$$

where $L = L(t, x, \dot{x})$ is the Lagrange function; $\dot{x} = \frac{dx}{dt}$.

Following [2], we write

$$L(t, x + \delta x, \dot{x} + \delta \dot{x}) = L(t, x, \dot{x}) + \delta L, \qquad (2)$$

where $\delta L = \left[\frac{d}{d\alpha} L(t, x + \alpha \delta x, \dot{x} + \alpha \delta \dot{x})\right]_{\alpha=0} = \frac{\partial L(t, x, \dot{x})}{\partial x}\delta x + \frac{\partial L(t, x, \dot{x})}{\partial \dot{x}}\delta \dot{x}$.

Substitute (2) into (1) and obtain the Jacobi equation.

$$\frac{\partial \delta L}{\partial x} - \frac{d}{dt}\frac{\partial \delta L}{\partial \dot{x}} = 0,$$

which is written as [2]

$$\left(L_{xx} - \frac{dL_{x\dot{x}}}{dt}\right)\delta x - \frac{d}{dt}\left(L_{\dot{x}\dot{x}}\delta \dot{x}\right) = 0. \qquad (3)$$

If $L = \frac{m\dot{x}^2}{2} - U(x,t)$, then the Euler–Lagrange and Jacobi equations have the form, respectively

$$U_x + m\ddot{x} = 0, \qquad (4)$$

$$U_{xx}\delta x + m\delta\ddot{x} = 0. \qquad (5)$$

It can be seen that using the Jacobi equation (5), closed systems of first-order differential equations can be constructed for three new dynamic variables.

$$\frac{d}{dt}\begin{pmatrix}\overline{\delta x^2}\\ \overline{\delta x \delta \dot{x}}\\ \overline{\delta \dot{x}^2}\end{pmatrix} = \begin{pmatrix} 0 & 2 & 0 \\ -\dfrac{U_{xx}}{m} & 0 & 1 \\ 0 & -\dfrac{2U_{xx}}{m} & 0 \end{pmatrix}\begin{pmatrix}\overline{\delta x^2}\\ \overline{\delta x \delta \dot{x}}\\ \overline{\delta \dot{x}^2}\end{pmatrix}, \qquad (6)$$

where $\overline{\delta x^2} = \overline{(\delta x)^2}$; $\overline{\delta \dot{x}^2} = \overline{(\delta \dot{x})^2}$; $\overline{\delta x \delta \dot{x}} = \overline{(\delta x \delta \dot{x})}$.

The recurrent formula has the form

$$\frac{d}{dt}\overline{\delta x^k \delta \dot{x}^s} = -s\frac{U_{xx}}{m}\overline{\delta x^{k+1}\delta \dot{x}^{s-1}} + k\overline{\delta x^{k-1}\delta \dot{x}^{s+1}}; \qquad (7)$$

$k + s = n.$

If $m\delta\dot{x} = \delta p_x$, then the system of equations (6) exactly coincides with the system of equations for the mean squares of the quantum fluctuations of the coordinate and momentum [3]

$$\begin{cases}\dfrac{d}{dt}\overline{\delta \hat{x}^2} = \dfrac{2}{m}\overline{\delta \hat{x}\delta \hat{p}_x}\\[4pt] \dfrac{d}{dt}\overline{\delta \hat{x}\delta \hat{p}_x} = \dfrac{1}{m}\overline{\delta \hat{p}_x^2} - U_{xx}\overline{\delta \hat{x}^2}\\[4pt] \dfrac{d}{dt}\overline{\delta \hat{p}_x^2} = -2U_{xx}\overline{\delta \hat{x}\delta \hat{p}_x},\end{cases} \qquad (8)$$

the initial conditions for which satisfy the Heisenberg uncertainty relations

$$\begin{aligned}(\overline{\delta \hat{x}\delta \hat{p}_x})^2\Big|_{t=0} &\le (\overline{\delta \hat{x}^2})\Big|_{t=0}\cdot(\overline{\delta \hat{p}_x^2})\Big|_{t=0},\\ (\overline{\delta \hat{x}^2})\Big|_{t=0}\cdot(\overline{\delta \hat{p}_x^2})\Big|_{t=0} &\ge \frac{\hbar^2}{4}.\end{aligned} \qquad (9)$$

The systems of differential equations (6-7) can be written as [4]

$$\frac{dW}{dt} = AW, \qquad (10)$$

where $W$ is the fundamental matrix of solutions, and $A$ is the matrix of coefficients for unknowns.

Using the Jacobi identity [4], it is possible to establish the relationship between the Wronsky determinant at the initial and final moments of time.

$$|W(t)| = |W(t_0)|\exp\left(\int_{t_0}^{t} SpA\, dt\right), \qquad (11)$$

where $SpA = 0$ is the sum of the diagonal elements of the fundamental matrices of differential equations (6-8) at $n = 2, 3, \ldots$ zero.

Thus, the ronsky determinants for the fundamental matrices of differential equations (6-8) are integrals of motion on the extremals of the Euler–Lagrange equation (4).

We construct the Liouville equation for the Jacobi equation (5)

$$\frac{\partial P}{\partial t} + \delta\dot{x}\frac{\partial P}{\partial \delta x} - \frac{U_{xx}}{m}\delta x\frac{\partial P}{\partial \delta\dot{x}} = 0. \qquad (12)$$

The solution to this equation is the distribution function of the fluctuations of the coordinate and velocity

$$P(\delta x, \delta\dot{x}, t) = \frac{1}{2\pi\cdot\Delta}\cdot\exp\left[-\frac{(\delta\dot{x})^2\overline{\delta x^2} + (\delta x)^2\overline{\delta\dot{x}^2} - 2(\delta x\delta\dot{x})\overline{\delta x\delta\dot{x}}}{2\Delta^2}\right], \qquad (13)$$

where $\Delta^2 = \overline{\delta x^2}\,\overline{\delta\dot{x}^2} - \overline{\delta x\delta\dot{x}}^2$, if the second moments $\overline{\delta x^2};\ \overline{\delta\dot{x}^2};\ \overline{\delta x\delta\dot{x}}$ satisfy the system of equations (6).

Using the system of equations (6), it can be shown that

$$\frac{d\Delta^2}{dt} = 0. \qquad (14)$$

The system of equations (9) exactly coincides with the system of equations for the mean squares of the quantum fluctuations of the coordinate and momentum [3,4]. Since it follows from formula (10) $\Delta^2(t) = const.$, for wave functions minimizing the Heisenberg uncertainty ratio, we obtain

$$\left(\overline{\delta\hat{x}^2}\right)\left(\overline{\delta\hat{p}_x^2}\right) - \left(\overline{\delta\hat{x}\delta\hat{p}_x}\right)^2 = \frac{\hbar^2}{4}. \qquad (15)$$

Thus, the distribution function (13) is localized in the vicinity of a classical trajectory, the size of which in the phase space is of the order of Planck's constant. Equation (15) determines the extremely small value of the emittance of the charged particle beam $\sim \hbar$, which is important in accelerator technology [5].